\documentclass[aps,prl,reprint,groupedaddress]{revtex4-2}
\pdfoutput=1
\usepackage{graphicx}
\usepackage{newtxtext}
\usepackage{newtxmath}
\usepackage[colorlinks=true, linkcolor=blue,anchorcolor=blue, citecolor=blue,urlcolor=blue]{hyperref}
\usepackage[normalem]{ulem}
\usepackage{amsmath}

\begin{document}
\title{Controllable chirality and band gap of quantum anomalous Hall insulators}

\author{Zhiming \surname{Xu}$^{1}$}
\author{Wenhui \surname{Duan}$^{1,2,3,4,5,6}$}
\email{duanw@tsinghua.edu.cn}
\author{Yong \surname{Xu}$^{1,2,3,4,7}$}
\email{yongxu@mail.tsinghua.edu.cn}
\affiliation{$^{1}$State Key Laboratory of Low Dimensional Quantum Physics and Department of Physics, Tsinghua University, Beijing 100084, China}
\affiliation{$^{2}$Tencent Quantum Laboratory, Tencent, Shenzhen, Guangdong 518057, China}
\affiliation{$^{3}$Frontier Science Center for Quantum Information, Beijing 100084, China}
\affiliation{$^{4}$Collaborative Innovation Center of Quantum Matter, Beijing 100084, China}
\affiliation{$^{5}$Institute for Advanced Study, Tsinghua University, Beijing 100084, China}
\affiliation{$^{6}$Beijing Academy of Quantum Information Sciences, Beijing 100193, China}
\affiliation{$^{7}$RIKEN Center for Emergent Matter Science (CEMS), Wako, Saitama 351-0198, Japan}

\begin{abstract}
Finding guiding principles to optimize properties of quantum anomalous Hall (QAH) insulators is of pivotal importance to fundamental science and applications. Here, we build a first-principles QAH material database of
chirality and band gap, explore microscopic mechanisms determining the QAH material properties, and obtain a general physical picture that can comprehensively understand the QAH data. Our results reveal that the usually neglected Coulomb exchange is unexpectedly strong in a large class of QAH materials, which is the key to resolve experimental puzzles. Moreover, we identify simple indicators for property evaluation and suggest material design strategies to control QAH chirality and gap by tuning cooperative or competing contributions via magnetic co-doping, heterostructuring, spin-orbit proximity, etc. The work is valuable to future research of magnetic topological physics and materials.
\end{abstract}

\maketitle

The quantum anomalous Hall (QAH) insulator (QAHI) is a novel two-dimensional (2D) magnetic topological phase that possesses Berry flux monopoles in the momentum space and displays quantized Hall conductance ($\sigma_{xy}$) robust against disorder \cite{haldane1988model,chang2013experimental}. QAHIs offer a versatile playground for exploring emerging quantum physics, such as dissipationless currents, topological magnetoelectric effects, Majorana fermions, etc~\cite{hasan2010colloquium,qi2011topological}. Despite intensive research effort, very few candidate materials are available experimentally and all of them must work at quite low temperatures~\cite{chang2013experimental,checkelsky2014trajectory,kou2014scale,chang2015high,mogi2015magnetic,li2019intrinsic,otrokov2019prediction,deng2020quantum,liu2020robust,ge2020high,liu2021magnetic,deng2021high,serlin2020intrinsic,li2021quantum}. Optimizing material properties of QAHI, especially the band gap, is of critical importance to experiment and applications, which demands in-depth research. For instance, the QAH gap of MnBi$_2$Te$_4$ predicted by theory is sizable and robust~\cite{li2019intrinsic,otrokov2019prediction}, whereas the gap detected by experiments is much smaller or even surprisingly vanishes~\cite{hao2019gapless,li2019dirac,chen2019toplogical}.

\begin{figure}[htb]
\includegraphics[width=\linewidth]{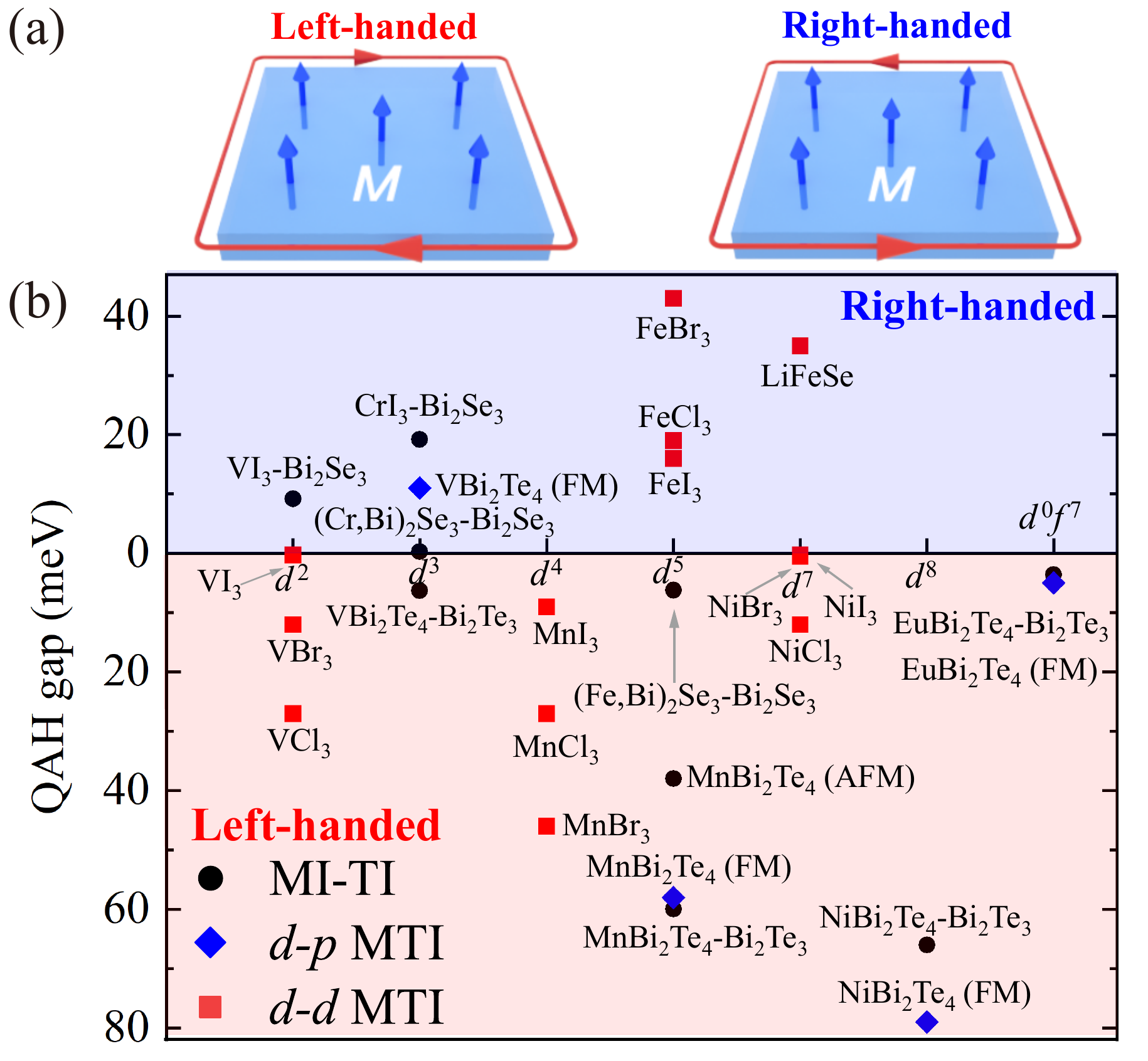}
\caption{\label{1} QAH chirality and material database. (a) Schematic QAHI with left-handed (left panel) or right-handed (right panel) chirality, denoted by the circulation orientation of chiral edge states with respect to magnetization {\bf M}. (b) First-principles database of QAH materials, where the QAH gap, chirality and magnetic orbital occupation are shown for three types of QAHIs.}
\end{figure}

Another significant but largely ignored property of QAHI is its chirality, which corresponds to the circulation direction of chiral edge modes [Fig. \ref{1}(a)]~\cite{qi2011topological}. In fact, the chirality is associated with the sign of QAH gap, which can be positive or negative, indicating the existence of compensating gap-opening mechanisms. Since the properties of chirality and band gap are intimately correlated to each other, the study of one property will help us optimize the other. However, the chirality property is usually neglected by first-principles studies. In experiment, QAH states with opposite chiralities have been discovered in Cr-doped (Bi,Sb)$_2$Te$_3$ and MnBi$_2$Te$_4$, showing opposite quantized $\sigma_{xy}$ for the same direction of magnetization~\cite{chang2013experimental,deng2020quantum,liu2020robust}, but the underlying reason remains as a puzzle. In this context, understanding the microscopic mechanism and finding guiding principles to control chirality and band gap of QAHIs become a key research subject of the field.

In this work, we systematically investigate physical mechanisms determining the QAH chirality and gap based on first-principles calculated material database and effective model Hamiltonian. We will first define the QAH chirality and correlate the chirality/gap with the sign/magnitude of Dirac mass, and then study microscopic mechanisms of Dirac mass generation induced by magnetic exchange interaction and spin-orbit coupling (SOC). By carefully designed first-principles experiments, we demonstrate that the $p$-$d$ Coulomb exchange is unusually strong in several QAH materials, which is essential to solve puzzling issues. Our work develops a general physical scenario to understand the QAH properties and provides simple indicators for property prediction, which is applicable to different classes of QAH materials. We also propose material design strategies to manipulate cooperative or competing Dirac-mass contributions for controlling QAH chirality and gap, which might find useful applications in research of QAH materials or magnetic topological materials in general.

According to experiments, we define the chirality of QAHIs as $\chi={\rm sgn}(\sigma_{xy}M_z)$, where $M_z$ is the out-of-plane $z$ component of magnetization, $\chi = +, -$ (or $\chi^+, \chi^-$) are called right-handed and left-handed, respectively. It is well known that QAHIs are characterized by the existence of chiral edge states and quantized $\sigma_{xy}$, and the propagation direction of chiral edge states is defined by ${\rm sgn} (\sigma_{xy})$. When $M_z$ is reversed, $\sigma_{xy}$ changes sign, giving the same $\chi$ (Fig. S1). Moreover, in contrast to the conventional chirality that relies on inversion symmetry breaking \cite{wagniere2007chirality}, the QAH phase can appear in inversion symmetric systems. Thus the QAH chirality is an intrinsic material property that is invariant under time reversal or spatial inversion operation.

First-principles study of QAH chirality is troubled by many sign ambiguities. For instance, $\sigma_{xy}$ is obtained from the TKNN formula~\cite{thouless1982quantized} by using different sign conventions: $\sigma_{xy}=\pm Ce^2/h$, where the Chern number $C=\pm \frac{1}{2\pi}\sum^{\rm occ.}_n\int_{\rm BZ} d^2{\textbf k} B_{n{\textbf k},z}$, the Berry curvature $\textbf{B}_{n{\textbf k}}= \pm \nabla_{{\textbf k}}\times\textbf{A}_{n{\textbf k}}$, the Berry connection  $\textbf{A}_{n{\textbf k}}= \pm i\langle u_{n{\textbf k}}|\nabla_{\textbf k}|u_{n{\textbf k}}\rangle$, and $|u_{n{\textbf k}}\rangle$ is the cell-periodic Bloch wavefunction of the $n^{\rm th}$ band. Moreover, the spin mentioned by theoretical studies may represent spin magnetic moment or spin angular momentum, which corresponds to opposite $M_z$. Possibly due to these ambiguities, a comprehensive theoretical study of QAH chirality is lacking. For simplicity we choose the plus sign for all the above formulas, assume $M_z > 0$, and use spin up to denote positive spin magnetic moment. Our convention implies $\chi={\rm sgn}(C)$.

Theoretically there are two fundamental models describing QAHIs, including the Haldane model~\cite{haldane1988model} and the Qi-Wu-Zhang (QWZ) model~\cite{qi2006topological}. We find that the QAH chirality and gap correspond to the sign and magnitude of Dirac mass, respectively, for both models (see the Supplemental Material (SM) Section I~\cite{SM}). Unfortunately, the Dirac mass as input parameter cannot be deduced from the models, whose properties should be investigated from a more fundamental level.

\begin{figure*}[htb]
\includegraphics[width=\linewidth]{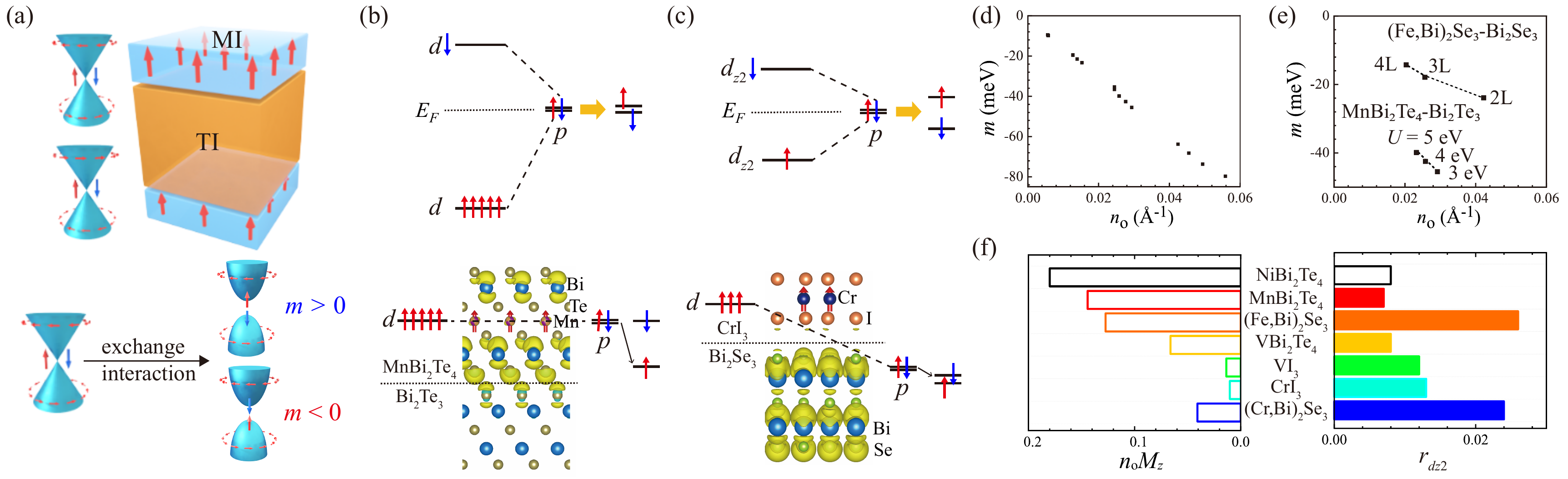}
\caption{\label{2}QAH states in MI-TI heterostructure. (a) Schematic illustration of MI-TI heterostructure and exchange induced Dirac mass with opposite signs. (b,c) Schematic diagrams of kinetic exchange (upper panels) and Coulomb exchange (lower panels) in (b) MnBi$_2$Te$_4$-Bi$_2$Te$_3$ and (c) CrI$_3$-Bi$_2$Se$_3$. Real-space distribution of TSSs is shown. (d,e) Relationship between Dirac mass ($m$) and $p$-$d$ orbital overlap ($n_{\rm o}$) in the high-spin $d^5$ systems obtained by artificially tuning (d) atom-resolved SOC strength, (e) Hubbard $U$ and TI film thickness. (f) $n_{\rm o}M_z$ and $r_{d_{z^2}}$ of MI-TI heterostructures for different MIs, which can be used to estimate $m_{\rm Coulomb}$ and $m_{\rm kinetic}$, respectively. Solid (hollow) bars denote positive (negative) contribution to $m$.}
\end{figure*}

\begin{figure}[htb]
\includegraphics[width=\linewidth]{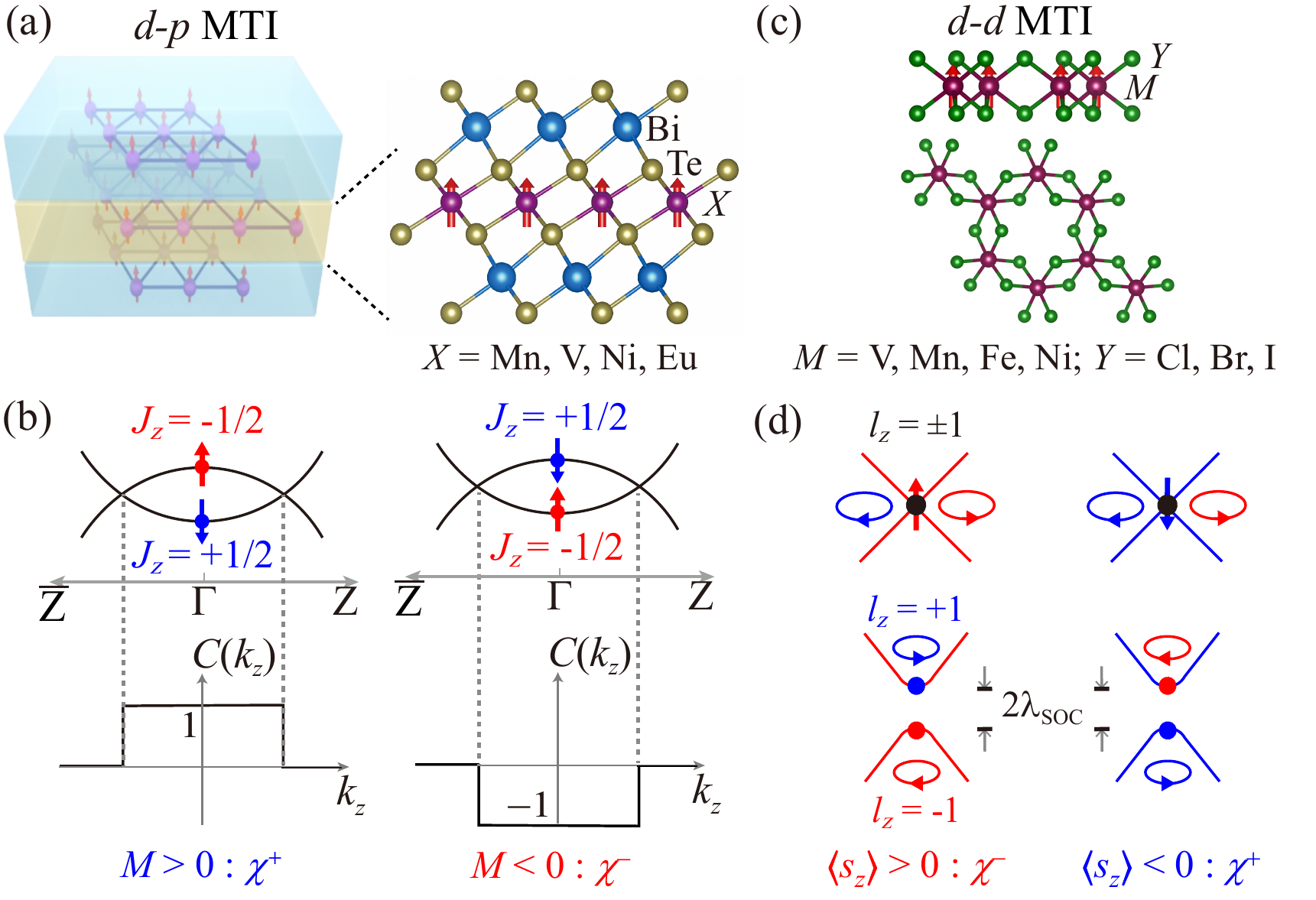}
\caption{\label{3}QAH states in intrinsic MTIs. (a,c) Schematic structures of $d$-$p$ MTI and $d$-$d$ MTI. (b) Chern number of 2D momentum space as a function of $k_z$ for topological Weyl semimetal. The quantum confinement creates type-II QAHI, whose chirality is defined by $J_z$ or spin splitting of topological bulk states near the Fermi level.  (d) Spin polarized Dirac bands of type-III QAHI, whose band degeneracy between states with pseudo-angular momenta $l_z=\pm1$ is lifted by SOC. The QAH chirality is determined by spin magnetization $\langle  s_z\rangle$.}
\end{figure}

We perform systematic first-principles calculations to build a QAH material database and summarize the calculation data in Fig. \ref{1}(b), which is a key result of this work. Detailed results are presented in the SM~\cite{SM}. The dependence of QAH gap and chirality on the magnetic $d$ orbital occupation is displayed. The QAH systems are classified into three types here, including magnetic insulator (MI)-TI heterostructure [type-I: MI-TI, Fig. \ref{2}(a)] featured by extrinsic magnetism, and intrinsic magnetic TI (MTI) with $d$-$p$ or $d$-$d$ magnetic-topological states characterized by different kinds of orbital combinations [type-II: $d$-$p$ MTI, Fig. \ref{3}(a); type-III: $d$-$d$ MTI, Fig. \ref{3}(c)]. Experimentally the QAH states observed in Cr-doped (Bi,Sb)$_2$Te$_3$ \cite{chang2013experimental} and MnBi$_2$Te$_4$ \cite{deng2020quantum,liu2020robust} are right- and left-handed, respectively. The chirality property can be reproduced by our calculations of Cr-alloyed Bi$_2$Se$_3$ and MnBi$_2$Te$_4$ flakes. It is generally believed that the QAH gap and chirality are intimately related to the $d$ orbital occupation. Such kind of relationship, if existing, would be subtle, considering that the same $d$ orbital occupation gives strongly fluctuated gap values and opposite chiralities [Fig. \ref{1}(b)]. Noticeably the QAH gap of MnBi$_2$Te$_4$- and NiBi$_2$Te$_4$-related systems can reach 60-80 meV, significantly larger than the others. These important features are to be understood.

Since the QAH gap is induced by the interplay of SOC and magnetism, the strength ratio $\uplambda_{\rm SOC} / \uplambda_{\rm ex}$ is a quantity critical to our study, where $\uplambda_{\rm SOC}$ denotes the SOC strength of topological states, and $\uplambda_{\rm ex}$ refers to the exchange interaction between magnetic and topological states. It is well established that the QAH physics is described by the low-energy effective Hamiltonian and the low-energy physics is majorly dictated by the weaker interaction. For type-I and type-II QAHIs, $\uplambda_{\rm SOC}$ is contributed by the strong SOC of heavy elements and $\uplambda_{\rm ex}$ by the relatively weak $p$-$d$ exchange interaction. In contrast, $\uplambda_{\rm ex}$ induced by Hund's coupling is usually much stronger than $\uplambda_{\rm SOC}$ for type-III QAHIs. Therefore, the low-energy physics is majorly defined by exchange interaction in the former systems and SOC in the latter. The classification of materials and properties is a key starting point of our study.

Figure \ref{2}(a) displays a schematic MI-TI heterostructure, which is composed of a TI flake in proximity with two MI layers on its top and bottom. For each side, topological surface states (TSSs) of TI are gapped by magnetic proximity, giving rise to the half quantum Hall effect~\cite{qi2011topological}. The two copies of gapped TSSs can result in an integer quantized Hall conductance. The study of low-energy topological physics (see the Supplementary Section I~\cite{SM}) suggests that the QAH gap is $2 (\vert m \vert-\vert\Delta\vert)$ and chirality $\chi={\rm sgn}(m)$, where $m$
is the Dirac mass of TSSs induced by exchange interaction and $\Delta$ denotes hybridization between top and bottom TSSs.

Exchange mechanisms between delocalized electrons and localized magnetic moments have been studied in the context of (diluted) magnetic semiconductors~\cite{larson1988theory,kacman2001spin,dietl2014dilute}. Generally there are two kinds of exchange interactions, including indirect hybridization-mediated kinetic exchange and direct Coulomb exchange, which are described by the Anderson Hamiltonian and Kondo Hamiltonian, respectively \cite{kacman2001spin}. Both of them introduce an effective Kondo-like Hamiltonian of the form $\mathbf{M} \cdot \mathbf{\sigma}$, by which the spin $\sigma$ of delocalized electrons is polarized by the magnetization $\mathbf{M}$ of localized magnetic moments. We apply the exchange theory to study the $p$-$d$ exchange-induced Dirac mass for TSSs. The Dirac mass has two types of contributions: $m = m_{\rm kinetic} + m_{\rm Coulomb}$, as classified by exchange mechanisms \cite{larson1988theory,kacman2001spin}. On kinetic exchange, the $p$-$d$ orbital hybridization is taken into account by a second-order perturbation: $m_{\rm kinetic} \propto -\sum_{d,s}{\rm sgn}(s \Delta^{s}_{pd}) t^2_{pd}/|\Delta^{s}_{pd}|$, where $t_{pd}$ is the hopping integral and $\Delta^{s}_{pd}$ denotes the energy cost of transferring one electron with spin $s$ from $p$ to $d$ orbitals. $m_{\rm kinetic}$ can be positive or negative, depending on the $d$-orbital occupation [Fig. S3(a)]. The direct Coulomb exchange has a first-order contribution: $m_{\rm Coulomb} \propto -\sum^{\rm occ.}_{d,s}{\rm sgn}(s)J_{pd}$, where $J_{pd}=\int \psi^*_p(\textbf{r}_1)\psi^*_d(\textbf{r}_2)\frac{1}{\textbf{r}_{12}}\psi_p(\textbf{r}_2)\psi_d(\textbf{r}_1)d^3r_1d^3r_2$ is the exchange integral \cite{xiang2013magnetic}. Since the dominated contribution comes from the majority spin ($s>0$) and $J_{pd} \ge 0$, the Coulomb exchange is always negative, which stabilizes parallel spins [Fig. S3(b)].

It is well established for (diluted) magnetic semiconductors that the $p$-$d$  Coulomb exchange can be safely neglected~\cite{larson1988theory,kacman2001spin,dietl2014dilute}, since $J_{pd}$ involves the $p$-$d$ overlap between different sites and is usually very small due to the localized nature of $d$ orbitals. The theory of kinetic exchange predicts right-handed chirality for the high-spin $d^5$ case (e.g. MnBi$_2$Te$_4$-Bi$_2$Te$_3$) [Fig. \ref{2}(b)], in contradiction with the experimental and theoretical data [Fig. \ref{1}(b)]. Moreover, in MnBi$_2$Te$_4$-related systems $\Delta_{pd}$ ($\Delta^{\uparrow}_{pd} \sim 4.7$ eV, $\Delta^{\downarrow}_{pd} \sim 2.5$ eV) is much larger than $t_{pd}$, implying very small $m_{\rm kinetic}$, in contradiction with the large magnitude of QAH gap ($\sim$60 meV). Furthermore, we apply the theory of kinetic exchange to study systems with $d^3$ occupation. Note that there exist $C_3$ rotation symmetries in these materials. The TSSs at the Dirac point are mainly composed of $p_z$, and $t_{pd}$ is mainly contributed by hybridization with $d_{z^2}$ as deduced from group theory and verified by first-principles calculations (Table S2). Herein the most relevant orbital is $d_{z^2}$ that only has spin up occupied, contributing positive $m_{\rm kinetic}$ [Fig. \ref{2}(c)]. This might explain $\chi^+$ of CrI$_3$-Bi$_2$Se$_3$ and (Cr,Bi)$_2$Se$_3$-Bi$_2$Se$_3$ but cannot explain $\chi^-$ of VBi$_2$Te$_4$-Bi$_2$Te$_3$. All these results indicate that the theory of kinetic exchange is incomplete to understand the properties.

It should be noted that in $X$Bi$_2$Te$_4$-Bi$_2$Te$_3$ ($X=$ Mn, V, Ni, or Eu) the TSSs near the Dirac point are delocalized within the $X$Bi$_2$Te$_4$ layer and show visible density distribution on the $d$-orbital sites [Fig. \ref{2}(b), Fig. S5]. This is in contrast to the usual case, such as in CrI$_3$-Bi$_2$Se$_3$ [Fig. \ref{2}(c)], which normally displays tiny $p$-$d$ overlap. We thus guess that the Coulomb exchange might not be negligible in the former systems. To verify the conjecture, we define and compute an overlap density to quantify the real-space $p$-$d$ orbital overlap: $n_{\rm o}= \int|\psi_p(x,y,z = z_d)|^2 dxdy$, which averages the density of TSSs $|\psi_p|^2$ over the plane $z=z_d$ at the $d$-orbital layer. In principle, a larger $n_{\rm o}$ would imply a bigger $J_{pd}$ and stronger Coulomb exchange. By focusing on the high-spin $d^5$ systems with weak kinetic exchange and artificially tuning $n_{\rm o}$ via varying computation parameters (e.g., SOC strength, Hubbard $U$ or TI film thickness), the relation between $m$ and $n_{\rm o}$ is systematically studied as shown in Figs. \ref{2}(d) and \ref{2}(e). Remarkably, our data clearly show a positive correlation between the two quantities, revealing important Coulomb exchange in MnBi$_2$Te$_4$-like materials.

Next we present a comprehensive picture to understand properties of type-I QAHIs. Since TSSs are gapped by kinetic exchange mainly via hybridizing with $d_{z^2}$, the ratio of $d_{z^2}$ ($r_{d_{z^2}}$) in the wavefunction of gapped TSSs could to some extent quantify the hybridization strength: $m_{\rm kinetic} \propto r_{d_{z^2}}$. The sign of $m_{\rm kinetic}$ is determined by the occupation of $d_{z^2}$, which is positive if only the spin up channel is occupied and is negative otherwise (Fig. S4). On Coulomb exchange, $m_{\rm Coulomb} \propto n_{\rm o}M_z$ and its sign is always negative. This provides a useful guiding principle to estimate the gap and chirality of type-I QAHIs by using the $d_{z^2}$ orbital occupation, $r_{d_{z^2}}$ and $n_{\rm o}M_z$ as indicators [Fig. \ref{2}(f)]. For instance, the following results are expected by the principle: (i) The large QAH gap of MnBi$_2$Te$_4$-Bi$_2$Te$_3$ is caused by the unusually strong Coulomb exchange. (ii) The Coulomb exchange is even stronger in NiBi$_2$Te$_4$-Bi$_2$Te$_3$, which together with a cooperative contribution of kinetic exchange result in an extremely large QAH gap. (iii) Both QAH systems show left-handed chirality as defined by Coulomb exchange. All type-I QAHIs of the database are understandable by the principle, as discussed in the SM~\cite{SM}.

The above guiding principle can be applied to study type-II QAHIs as well. As shown in Fig. \ref{3}(a) multilayer $X$Bi$_2$Te$_4$ ($X=$ Mn, V, Ni, or Eu) in the ferromagnetic (FM) configuration are studied. Their topological properties are interpreted via the bulk phase. The 3D FM bulk is topological Weyl semimetal (WSM) for MnBi$_2$Te$_4$ \cite{li2019intrinsic} and NiBi$_2$Te$_4$, and high-order MTI \cite{li2020tunable} for VBi$_2$Te$_4$ and EuBi$_2$Te$_4$ (Fig. S6). By quantum confinement, their multilayers become QAHIs, whose $C$ grows with layer thickness for the WSM~\cite{ge2020high} or $C = 1$ for the high-order MTI~\cite{li2020tunable}. These QAH states are described by the QWZ model~\cite{qi2006topological} with $\chi={\rm sgn}(M)$, where $M$ denotes the Dirac mass (see the Supplementary Section I). Here the relevant $p$ orbital is topological bulk states (TBSs) near the Fermi level, corresponding to $p_z$ of Te (Bi/Te) for the WSM (high-order MTI) phase. With $C_3$ rotation symmetry, the pseudo-angular momentum $J_z=\pm1/2$ is a good quantum number at the $\Gamma$ point. ${\rm sgn}(M)$ refers to the band order between the $J_z=+1/2$ and $J_z=- 1/2$ states, whose spin component is mainly spin-down and spin-up, respectively [Fig. \ref{3}(b)]. In this sense, $\chi$ is related to the spin splitting of TBSs, which is majorly dictated by $p$-$d$ exchange interaction. Therefore, properties of QAHIs can be understood by applying the above guiding principle to TBSs (instead of TSSs), as described in detail in the SM~\cite{SM}.

Take multilayer MnBi$_2$Te$_4$ for example. Kinetic exchange introduces small, positive Dirac mass, whereas Coulomb exchange induces compensating contribution of unusually large magnitude, similar as found for MnBi$_2$Te$_4$-Bi$_2$Te$_3$. In addition to using $d_{z^2}$ projection and $n_{\rm o}M_z$ as indicators, we also applied a method of Wannier tight-binding Hamiltonian analysis to quantify strength of kinetic and Coulomb exchanges~\cite{SM}. Our results reveal that strong Coulomb exchange is prevalent in MnBi$_2$Te$_4$-family materials, 
distinct from conventional material systems. This is caused by the delocalized nature of TBSs (evidenced by noticeable density distribution near magnetic elements shown in Fig. S9), possibly related to the unique material features, such as the small electronegativity difference between anions and cations~\cite{li2020tunable}, the topological band inversion that enhances anion-cation orbital hybridization, etc. Previously we discovered extremely long-range superexchange in the materials~\cite{li2020tunable}, which might share the same origin.

Type-III QAHIs are understood in a different scenario. Figure \ref{3}(c) displays a few candidate materials, including monolayer transition metal trihalides $MY_3$ ($M$ = transition metal element, $Y$ = halogen element) ~\cite{he2017near,huang2017quantum,sheng2017monolayer,wang2018high,li2018electrically,sui2020model}. As illustrated in Fig. \ref{3}(d), the low-energy topological physics is described by a Haldane-like model (see the Supplementary Section I~\cite{SM}) with an SOC-induced Dirac mass $m=- \uplambda_{\rm SOC} \langle  s_z\rangle$, where $\uplambda_{\rm SOC}$ denotes the effective SOC of transition metal $d$ orbitals and $\langle  s_z\rangle$ is the  expectation value of out-of-plane spin component of Dirac bands. The QAH gap is $2|m|$ and chirality $\chi={\rm sgn}(m)$. Typically $\uplambda_{\rm SOC}$ is positive, giving $\chi = -{\rm sgn} (\langle  s_z\rangle)$. This rule applies to the first-principles data (see Table S9 \cite{SM}). Note that in rare cases the effective $\uplambda_{\rm SOC}$ might get negative, which would reverse the QAH chirality.

\begin{figure}[htb]
\includegraphics[width=\linewidth]{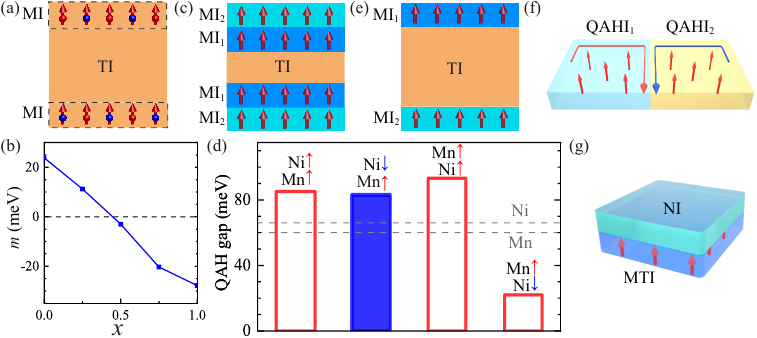}
\caption{\label{4}QAH material design. (a) Magnetic co-doping. (b) Exchange-induced Dirac mass as a function of Fe-Cr element ratio ($x$) for (Cr$_{1-x}$Fe$_x$,Bi)$_2$Te$_3$-Bi$_2$Te$_3$ heterostructure. (c,e) Vertical and (f) in-plane magnetic heterostructuring. (d) QAH gap and chirality ($\chi^+$: solid bars; $\chi^-$: hollow bars) of heterostructures displayed in (c). Results are shown for different combinations of MI$_1$ and MI$_2$ as well as for normal MI-TI heterostructures (MI = $X$Bi$_2$Te$_4$, $X$ = Mn or Ni; TI = Bi$_2$Te$_3$). In (f) chiral boundary states appear at the FM domain wall between QAHIs with opposite chiralities. (g) Spin-orbit proximity. A MTI is in proximity with a normal insulator (NI).}
\end{figure}

So far we have developed a general physical picture to understand QAH material properties from the microscopic level. The scenario not only works for the whole first-principles QAH database but also helps resolving puzzles of QAH experiments mentioned in the introduction (see the Supplementary Section IV~\cite{SM}). As applications, we suggest some material design strategies to control QAH chirality and band gap by optimizing cooperative or competing Dirac-mass contributions. This could be realized, for instance, by tuning the balance between different Dirac-gap opening mechanisms or by building heterostructures. We list a few examples here. (i) {\it Magnetic co-doping} [Fig. \ref{4}(a)]: Our study of (Fe, Cr) doped Bi$_2$Te$_3$ shows that the exchange-induced Dirac mass can be tuned from 23.8 meV to -27.8 meV by changing the Fe-Cr element ratio [Fig. \ref{4}(b)]. (ii)  {\it Magnetic heterostructuring}: (a) The idea is to manipulate exchange interaction by coupling two magnetic layers (MI$_1$ and MI$_2$) together [Fig. \ref{4}(c))]. Our study of heterostructures MI$_2$/MI$_1$/TI/MI$_1$/MI$_2$ (TI = Bi$_2$Te$_3$; MI$_1$, MI$_2$ = monolayer $X$Bi$_2$Te$_4$ with up or down spin magnetization) demonstrates an effective control of QAH gap and chirality by heterostructuring as well as rich topological phase transitions induced by varying magnetic configuration [Fig. \ref{4}(d) and Table S10]. (b) Our study of heterostructures MI$_1$/TI/MI$_2$ [Fig. \ref{4}(e)] shows that novel quantum states of matter, including antiferromagnetic QAHI and FM axion insulator, emerge when the exchange-induced Dirac mass is opposite between the two magnetic layers (Fig. S15). (c) 2D in-plane heterostructures QAHI$_1$/QAHI$_2$ could be built with two QAHIs of opposite chiralities [Fig. \ref{4}(f)]. Thus dissipationless chiral conduction channels can be patterned at the FM domain walls, which are more easily controlled than antiferromagnetic cases via magnetic field, useful for low-power integrated circuits. (iii) {\it Spin-orbit proximity}: The idea is to enhance the strength of SOC by proximity with heavy-element materials [Fig. \ref{4}(g)], so as to increase the QAH gap. This is demonstrated by our study of monolayer MnI$_3$ in proximity with Bi$_2$Se$_3$ (Fig. S16). In fact, the above strategies could be applied simultaneously, which provides great opportunities for optimizing QAH material properties. In brief, we believe that the physical mechanisms and property optimizing strategies are not only useful for the QAH research, but could also find general applications in future study of magnetic topological materials.

\begin{acknowledgments}
This work was supported by the Basic Science Center Project of NSFC (Grant No. 51788104), the National Science Fund for Distinguished Young Scholars (Grant No. 12025405), the National Natural Science Foundation of China (Grant No. 11874035), the Ministry of Science and Technology of China (Grant Nos. 2018YFA0307100 and 2018YFA0305603), the Beijing Advanced Innovation Center for Future Chip (ICFC), and the Beijing Advanced Innovation Center for Materials Genome Engineering.
\end{acknowledgments}

%

\end{document}